\documentclass[11pt,a4paper,oneside]{article}
\usepackage{lineno,hyperref}

\usepackage{graphicx}              
\usepackage{epsfig}
\usepackage{amsmath}               
\usepackage{amssymb}
\usepackage{epstopdf}
\usepackage{verbatim}
\usepackage{mathptmx}
\usepackage{mathalpha}
\usepackage{mathtools}
\usepackage[scale=0.80]{geometry}
\usepackage{graphicx, times}
\usepackage{amsfonts}              
\usepackage{amsthm}                
\usepackage{multicol}
\usepackage{algorithm}
\usepackage{algorithmic}
\usepackage{varwidth}
\usepackage{parskip}
\usepackage{hyperref}
\usepackage{rotating}
\usepackage[numbers,sort&compress]{natbib}
\usepackage{multirow}
\usepackage{pdflscape}
\usepackage[numbers]{natbib}
\usepackage{caption}
\usepackage{xcolor}
\usepackage{color}
\usepackage{comment}
\usepackage{subcaption}
\newcommand*{\rom}[1]{\expandafter\@\romannumeral #1}

\newcommand{\bea}{\begin{eqnarray}}
	\newcommand{\eea}{\end{eqnarray}}
\newcommand{\bee}{\begin{eqnarray*}}
	\newcommand{\eee}{\end{eqnarray*}}

\begin{document}
\author{Khomesh R. Patle$^{}$\footnote{khomeshpatle5@gmail.com}, G. P. Singh$^{}$\footnote{gpsingh@mth.vnit.ac.in}
\vspace{.3cm}\\
${}^{}$ Department of Mathematics,\\ Visvesvaraya National Institute of Technology, Nagpur, 440010, Maharashtra, India.
\vspace{.3cm}
\date{}}
\title{Transiently accelerating cosmological model with Gong-Zhang parametrization in $f(T)$ teleparallel gravity}
\maketitle
\begin{abstract} \noindent
We present the cosmic expansion scenario in the framework of $f(T)$ gravity by employing a dark energy equation of state (EoS) parameter. Specifically, we proceed with the power-law form of the function $f(T) = \alpha$$(-T)^{n}$, in conjunction with the Gong-Zhang parametrization of the dark energy EoS. We derive the expansion rate in terms of the redshift for the considered model, providing deeper insights into the underlying cosmic dynamics. The model is further utilized to explore the expansion history of the universe and the evolution of several cosmological parameters. By using the Bayesian methods based on the $\chi^{2}$-minimization technique, the median values of the model parameters are determined for the cosmic chronometer (CC) and joint (CC+Pantheon) datasets. The evolution of the deceleration parameter, energy density, pressure, EoS parameter and the energy conditions for dark energy is analyzed in detail. The model captures the observed acceleration as a transient phenomenon, followed by future deceleration. Additionally, the nature of both geometrical and dynamical diagnostics robustly indicates a quintessence-like behavior at the present epoch. Finally, the thermodynamic viability of the model is confirmed through the generalized second law of thermodynamics and the estimated age of the universe further supports the model’s compatibility with astronomical observations.
\end{abstract}

\section{Introduction}\label{sec:1}
A wide range of independent cosmological observations~\cite{1998AJ....116.1009R,1999ApJ...517..565P,2020A&A...641A...6P} have firmly established that the universe is currently undergoing an accelerated phase of expansion. This groundbreaking discovery has led to the hypothesis of dark energy (DE), a dominant component believed to be responsible for driving this late-time acceleration. Despite its substantial contribution to the total energy density of the universe, the fundamental nature of DE remains a mysterious problem in modern cosmology. Observational evidence indicates that dark energy together with dark matter, constitutes approximately $95$ – $96 \%$ of the total cosmic energy content~\cite{weinberg1989cosmological}. Among the various candidates proposed for DE, the cosmological constant ($\Lambda$) stands out as the simplest and most successful model. The standard $\Lambda$CDM framework provides an excellent agreement with current observational data. However, it suffers from two major theoretical challenges, namely the fine-tuning problem and the coincidence problem~\cite{di2021realm,carroll2001cosmological,padmanabhan2003cosmological,copeland2006dynamics}. These longstanding issues have motivated the exploration of alternative approaches to explain the observed cosmic acceleration. In this regard, researchers have investigated a broad class of models, including those involving exotic forms of matter and energy as well as modifications to Einstein’s theory of General Relativity (GR). In particular, modified gravity theories have emerged as compelling alternatives, offering the possibility of explaining the accelerated expansion of the universe without the need to introduce an explicit DE component. Consequently, a wide variety of modified gravity models have been proposed and extensively studied in the literature~\cite{buchdahl1970non,harko2011f,nojiri2011unified,jimenez2018coincident,nojiri2017modified,bamba2010finite,elizalde2010lambdacdm,harko2010f,capozziello2019extended,capozziello2023role,kotambkar2017anisotropic,lalke2023late,hulke2020variable,garg2024cosmological,singh2025observational,singh2024conservative,chaudhary2025extracting,goswami2024flrw,shukla2025multi,escamilla2024exploring,patle2026dynamical}.
\vspace{0.2cm}\\
One of the most important and extensively studied modifications of gravity is $f(T)$ teleparallel gravity~\cite{Bengochea,cai2016f}. This framework generalizes the teleparallel equivalent of GR (TEGR) by promoting the torsion scalar $T$ in the gravitational action to an arbitrary function $f(T)$. In this sense, it bears a strong conceptual resemblance to $f(R)$ gravity, where the Einstein-Hilbert Lagrangian is extended to a function of the Ricci scalar ($R$). The key distinction between these theories, lies in their underlying geometric formulation. While GR is constructed on the torsion-free Levi-Civita connection, interpreting gravity as a manifestation of spacetime curvature, teleparallel gravity is based on the curvature-free Weitzenböck connection, attributing gravitational effects to torsion instead. At the level of TEGR, this alternative geometric description is dynamically equivalent to GR. However, its extension to $f(T)$ gravity gives rise to modified field equations, leading to a rich and diverse set of cosmological implications. Due to its mathematical simplicity and its ability to account for the late-time cosmic dynamics of the universe without the need for an explicit DE component, $f(T)$ gravity has attracted considerable attention in recent years. 
\vspace{0.2cm}\\
A broad range of studies have explored its cosmological viability from different perspectives. These include investigations of dynamical cosmological solutions describing the evolution of the universe~\cite{paliathanasis2016cosmological} as well as analyses of thermodynamic properties within the teleparallel framework~\cite{salako2013lambdacdm}. Cosmographic techniques have been employed to reconstruct the expansion history in a model-independent manner~\cite{capozziello2011cosmography} while classical energy conditions have been examined to assess the physical plausibility of various $f(T)$ models~\cite{liu2012energy}. In addition, alternative early-universe scenarios, such as matter-bounce cosmologies, have also been formulated within this context~\cite{cai2011matter}. Observational studies further reinforce the significance of $f(T)$ gravity. For instance, Zhadyranova et al.~\cite{zhadyranova2024exploring} conducted a detailed investigation of late-time cosmic acceleration within the framework of a linear $f(T)$ model, employing observational data to constrain the model parameters. A comprehensive and systematic overview of the theoretical structure and cosmological implications of $f(T)$ gravity is provided in Ref.~\cite{cai2016f}. Among other significant developments, Bamba et al.~\cite{bamba2011equation} investigated the behavior of the DE EoS parameter ($\omega_{DE}$) for exponential, logarithmic and hybrid forms of $f(T)$. Similarly, Paliathanasis et al.~\cite{paliathanasis2014new} carried out a systematic Noether symmetry analysis, shedding light on the conserved quantities and integrability features of the theory. In addition, Capozziello et al.~\cite{capozziello2017model} introduced a model-independent numerical scheme for addressing the modified Friedmann equations in the context of teleparallel cosmology. Taken together, these investigations along with numerous other studies reported in the literature~\cite{shekh2025cosmographical,duchaniya2024attractor,maurya2023anisotropic,maurya2022accelerating,bamba2016bounce,bhar2024anisotropic,nunes2016new,chaudhary2024constraints,duchaniya2022dynamical,chakraborty2023classical,maurya2024role,dixit2021probe,bahamonde2023teleparallel,patle2026revisiting,das2023study,ren2022gaussian}, highlight the broad theoretical richness and strong observational relevance of $f(T)$ gravity, establishing it as a promising and robust framework for addressing some of the fundamental challenges in modern cosmology.
\vspace{0.2cm}\\
The primary objective of this work is to delve into the extended $f(T)$ modified gravity, commencing from the TEGR rather than the curvature-based formulation of GR. In this study, we adopt the Gong-Zhang parametrization for the EoS parameter and derive exact solutions of the modified Friedmann equations within the homogeneous and isotropic Friedmann-Lemaître-Robertson-Walker (FLRW) spacetime. In particular, we consider a power-law form of $f(T)$ model given by $f(T)$ = $\alpha$$(-T)^{n}$, where $\alpha$ and $n$ are free model parameters. These parameters are subsequently constrained using a comprehensive set of observational datasets, including the cosmic chronometer (CC) dataset and the joint (CC+Pantheon) dataset. This research work focuses on a detailed investigation of the late-time cosmic dynamics of the universe within the $f(T)$ gravity framework by examining the evolution of various cosmological parameters.  
\vspace{0.2cm}\\
This work has been divided into six sections: Section~(\ref{sec:2}) presents the fundamental formalism of $f(T)$ gravity and the corresponding field equations within the spatially flat FLRW framework, thereby establishing the theoretical foundation of the present study. Section~(\ref{sec:3}) examines the background cosmological dynamics, providing a comprehensive account of the expansion history in the considered model. Section~(\ref{sec:4}) is devoted to constraining the model parameters through Bayesian statistical inference, employing cosmic chronometer (CC) and joint (CC+Pantheon) observational datasets. The physical and dynamical properties of the model are investigated in Section~(\ref{sec:5}), where we systematically analyze the evolution of key cosmological quantities, including the deceleration parameter, energy density, pressure and the EoS parameter for DE. Furthermore, we explore the validity of energy conditions, assess geometrical and dynamical diagnostics, examine the thermodynamic behavior and estimate the age of the universe. Finally, Section~(\ref{sec:6}) summarizes the principal results and presents the concluding remarks.
\section{The $f(T)$ gravity framework and field equations}\label{sec:2}
The $f(T)$ theory of gravity represents a modification of the gravitational framework constructed on the torsion scalar $T$, in which the gravitational action is generalized to an arbitrary function of torsion. Similar to TG, this theory is formulated using orthonormal tetrad fields defined on the tangent space at each point of the spacetime manifold, providing an alternative geometric description of gravity. Within this formalism, the geometry of spacetime is fully characterized by these tetrad fields. In general, the spacetime line element can be expressed as
\begin{equation}{\label{1}}
ds^{2}= g_{\mu\nu}dx^{\mu}dx^{\nu}= \eta_{lm}\theta^{l}\theta^{m},
\end{equation}
with the components
\begin{equation}{\label{2}}
dx^{\mu} = e^{\mu}_{l}\theta^{l},~~~~~  \theta^{l}= e^{l}_{\mu} dx^{\mu},
\end{equation}
where $\eta_{lm}$= diag$(-1,1,1,1)$ is the metric associated with flat spacetime, and $\left\{e^{l}_{\mu}\right\}$ are the components the tetrad. These tetrads satisfy the conditions
\begin{equation}{\label{3}}
e^{~~\mu}_{l} e^{l}_{~~\nu}= \delta^{\mu}_{\nu},~~~~~  e^{~~l}_{\mu} e^{\mu}_{~~m}= \delta^{l}_{m}.
\end{equation}
In $f(T)$ gravity, the underlying geometric structure is based on the Weitzenböck connection~\cite{aldrovandi2012teleparallel}, defined as
\begin{equation}{\label{4}}
	\Gamma^{\alpha}_{\mu \nu}= e_{l}^{~\alpha} \partial_{\mu} e^{l}_{~\nu}= -e^{l}_{~\mu}\partial_{\nu} e^{~\alpha}_{l}.
\end{equation}
Based on this connection, the components of the torsion tensor can be written as~\cite{linder2010einstein}
\begin{equation}{\label{5}}
T^{\alpha}_{~\mu \nu} = -\left(\Gamma^{\alpha}_{\nu \mu}-\Gamma^{\alpha}_{\mu \nu}\right)= - e^{~\alpha}_{l} \left(\partial_{\mu} e^{l}_{~\nu} - \partial_{\nu} e^{l}_{~\mu}\right).
\end{equation}
This tensor plays a key role in defining the contorsion tensor, which is given by
\begin{equation}{\label{6}}
K^{\mu \nu}_{~\alpha} = -\frac{1}{2} \left(T^{\mu \nu}_{~\alpha} - T^{\nu \mu}_{\alpha} - T^{~\mu \nu}_{\alpha}\right),
\end{equation}
which, together with the torsion tensor, leads to the definition of the tensor
\begin{equation}{\label{7}}
S^{~\mu \nu}_{\alpha} = \frac{1}{2} \left(K^{\mu \nu}_{\alpha} + \delta^{\mu}_{\alpha} T^{\lambda \nu}_{~\lambda} - \delta^{\nu}_{\alpha}T^{\lambda \mu}_{~\lambda}\right).
\end{equation}
The torsion scalar $T$ is defined as a scalar quantity constructed from the torsion tensor and the superpotential $S^{~\mu \nu}_{\alpha}$, and is given by~\cite{cai2016f,maluf2013teleparallel}
\begin{equation}{\label{8}}
T= S^{~\mu \nu}_{\alpha} T^{\alpha}_{~\mu \nu} = \frac{1}{2}T^{\alpha \mu \nu } T_{\alpha \mu \nu} + \frac{1}{2}T^{\alpha \mu \nu } T_{\nu \mu \alpha} - T^{~\alpha}_{\alpha \mu } T^{\nu \mu}_{~\nu}.
\end{equation} 
Further, the action corresponding to this gravitational theory is given by~\cite{Bengochea}
\begin{equation}{\label{9}}  
	S=  \frac{1}{2\kappa^{2}}\int d^{4}xe \left[T+f(T)\right] + \int d^{4}xe L_{m},
\end{equation}
here, $e$ denotes the determinant of the tetrad, defined as $e= det (e^{l}_{~\mu}) = \sqrt{-g}$. By varying the action (\ref{9}) with respect to the tetrad fields, one obtains the field equations of $f(T)$ gravity:
\begin{equation}{\label{10}}
S^{~\nu \rho}_{\mu} \partial_{\rho} T f_{TT} + [e^{-1} e^{l}_{\mu} \partial_{\rho} (ee^{~\mu}_{l} S^{~\nu \lambda}_{\alpha}) + T^{\alpha}_{~\lambda \mu} S^{~\nu \lambda}_{\alpha}] f_{T} + \frac{1}{4}\delta^{\nu}_{\mu} f = \frac{\kappa^{2}}{2} \mathit{T}_{\mu}^{\nu},
\end{equation}
where $f_{T}= \frac{\partial f}{\partial T}$, $f_{TT}= \frac{\partial^{2} f}{\partial T^{2}}$ and $\mathit{T}_{\mu}^{\nu}$ denotes the energy-momentum tensor, defined as
\begin{equation}{\label{11}}
	\mathit{T}_{\mu}^{\nu} = (\rho + \mathit{p}) u_{\mu} u^{\nu} + p \delta^{\nu}_{\mu},
\end{equation}
where $p$ and $\rho$ correspond to the pressure and energy density, respectively, of the matter content of the universe. The associated four-velocity field of this ordinary matter, $u^{\mu}$, satisfies the normalization condition $u^{\mu} u_{\nu}=-1$.
\vspace{0.1cm}\\
In this study, we consider a spatially flat FLRW metric, which is widely adopted for implementing the aforementioned theory within a cosmological framework. This assumption enables the derivation of the modified Friedmann equations. The line element corresponding to the flat FLRW spacetime is given by~\cite{linder2010einstein}
\begin{equation}{\label{12}}
	ds^{2}=-dt^{2}+a^{2}(t) \delta_{lm} dx^{l} dx^{m},
\end{equation}
where $a(t)$ represents the scale factor. For the line element (\ref{12}), the corresponding torsion scalar takes the form $T=-6H^{2}$.
\vspace{.1cm}\\
The Friedmann equations corresponding to the metric (\ref{12}) are given by~\cite{cai2016f}:
\begin{equation}{\label{13}}
6H^{2}+ 12H^{2}f_{T}+f = 2 \kappa^{2}\rho,
\end{equation}
\begin{equation}{\label{14}}
2\left(2\dot{H}+3H^{2}\right)+f+4 \left(\dot{H}+3H^{2}\right) f_{T}-48H^{2}\dot{H}f_{TT}=-2 \kappa^{2}p.
\end{equation}
Here, an overdot denotes differentiation with respect to the cosmic time $t$ and $H$ represents the Hubble parameter. The quantities $\rho$ and $p$ correspond to the energy density and pressure of the matter content, respectively. By setting $\kappa^{2}=1$, equations (\ref{13}) and (\ref{14}) can be expressed as
\begin{equation}{\label{15}}
	3H^{2}= \rho + \rho_{DE},
\end{equation}
\begin{equation}{\label{16}}
-2 \dot{H}-3H^{2}= p+p_{DE}.
\end{equation}
Here, the energy density and pressure attributed to DE are specified as follows
\begin{equation}{\label{17}}
	\rho_{DE}= -6H^{2}f_{T}-\frac{1}{2}f,
\end{equation}
\begin{equation}{\label{18}}
	p_{DE}= \frac{1}{2}f + 2\left(\dot{H}+3H^{2}\right) f_{T}-24\dot{H} H^{2} f_{TT}.
\end{equation}
Using equations~(\ref{17}) and (\ref{18}), the EoS parameter of DE is obtained as
\begin{equation}{\label{19}}
\omega_{DE}=\frac{p_{DE}}{\rho_{DE}} = -1-\frac{4\dot{H} \left(f_{T}-12H^{2}f_{TT}\right)}{f+ 12H^{2}f_{T}}.
\end{equation}
\section{Background dynamics in $f(T)$ cosmology}\label{sec:3}
Within the framework of $f(T)$ modified gravity, power-law models have been widely studied for reconstructing and describing various evolutionary scenarios of the universe. For instance, Karami and Abdolmaleki~\cite{karami2012generalized} examined the validity of the generalized second law of thermodynamics in $f(T)$ gravity by considering viable models, including the power-law form. Rezazadeh et al.~\cite{rezazadeh2016power} explored both power-law and intermediate inflationary scenarios within this framework, highlighting the role of $f(T)$ gravity in early-universe cosmology. Furthermore, Basilakos~\cite{basilakos2016linear} investigated the linear growth of cosmic structures in the context of power-law $f(T)$ gravity, while Malekjani et al.~\cite{malekjani2017spherical} analyzed the spherical collapse model and cluster number counts within the same framework. Boko and Houndjo~\cite{boko2020cosmological} studied cosmological models with viscous fluids, focusing on the occurrence of finite-time singularities in power-law $f(T)$ gravity. Recently, Kumar et al.~\cite{kumar2023new} provided updated observational constraints on $f(T)$ gravity using combined Planck-CMB and SNeIa datasets. The study offers important insights into its viability as an alternative theory at cosmological scales. Motivated by these studies, in this work we examine the late-time behavior of the universe by considering the Gong-Zhang parametrization of the EoS parameter within the framework of the following power-law $f(T)$ model as 
\begin{equation}{\label{20}}
f(T) = \alpha (-T)^{n},
\end{equation}
where $\alpha$ and $n$ are model parameters. The primary objective of this study is to investigate the dynamical evolution of the universe and the properties of DE using observational data. A wide range of parametrizations have been proposed in the literature, many of which involve two or more free parameters. For instance, Mandal et al.~\cite{mandal2023cosmological} examined cosmological constraints on the power-law $f(Q)$ modified gravity model using the CPL parametrization of the DE EoS. Similarly, Arora et al.~\cite{arora2021constraining} analyzed constraints on the effective EoS parameter within the framework of $f(Q, T)$ gravity, focusing on the behavior of DE and its cosmological implications. Koussour and De~\cite{koussour2023observational} studied observational constraints on cosmological models in $f(Q)$ gravity using the EoS parametrization as $\omega(z)=(-1-3\beta(1+z)^{3})^{-1}$. In a related context, Myrzakulov et al.~\cite{myrzakulov2023constrained} investigated a non-linear $f(R, L_{m})$ DE model employing the same EoS form and performed a Bayesian analysis using cosmic chronometer and Pantheon SNeIa datasets to probe the evolution of DE. However, the use of such parametrizations to investigate the cosmic dynamics can be intricate, as it involves introducing additional cosmological parameters, namely $H_{0}$ and $n$, to the model. For this reason, it is preferable to use a parametrization with a single free parameter. In light of this consideration, we adopt the Gong-Zhang parametrization of the EoS for DE~\cite{gong2005probing}:
\begin{equation}{\label{21}}
\omega_{DE}(z)=\frac{\omega_{0}}{1+z} \exp\left(\frac{z}{1+z}\right),
\end{equation}
where $\omega_{0}$ denotes the present-day value of the DE EoS. In the limit $z \gg 1$ (early universe), $\omega_{DE}(z) \to 0$, indicating that DE behaves like pressureless matter in the past. At the present epoch ($z = 0$), $\omega_{DE}(0) = \omega_0$ denotes the current value of the DE EoS. Importantly, as $z \to -1$ (in the future), $\omega_{DE}(z)$ again approaches zero.
\vspace{0.1cm}\\
Using equations~(\ref{19}), (\ref{20}) and (\ref{21}), we obtain
\begin{equation}{\label{22}}
\dot{H}+\frac{3H^{2}}{2n}+\frac{3\omega_{0}H^{2}}{2n(1+z)}\exp\left(\frac{z}{1+z}\right)=0.
\end{equation}
Employing the relation $\frac{1}{H}\frac{d}{dt} = -(1+z)\frac{d}{dz}$, the given equation can be recast as a first-order differential equation,
\begin{equation}{\label{23}}
\frac{d\ln (H)}{dz}-\frac{3}{2n(1+z)}-\frac{3\omega_{0}}{2n(1+z)^{2}}\exp\left(\frac{z}{1+z}\right)=0.
\end{equation}
By integrating the aforementioned equation, we derive the Hubble parameter in terms of redshift as
\begin{equation}{\label{24}}
H(z)=H_{0} (1+z)^{\frac{3}{2n}} \exp\left(\frac{3\omega_{0}}{2n}\left[\left(\exp\left(\frac{z}{1+z}\right)\right)-1\right]\right),
\end{equation} 
where $H_{0}$ defined as Hubble parameter value at $z=0$. The model possesses three parameters $H_{0}$, $\omega_{0}$ and $n$.
\section{Observational constraints}\label{sec:4}
In this section, we utilize a Bayesian inference framework to examine the compatibility of the proposed cosmological model with observational datasets. The key model parameters $H_{0}$, $\omega_{0}$ and $n$, appearing in the parameterized form of the Hubble parameter in equation~(\ref{24}), are constrained using the cosmic chronometer (CC) dataset along with the combined (CC+Pantheon) dataset. The estimation of these parameters is performed by minimizing the $\chi^{2}$ function, together with the Markov chain Monte Carlo (MCMC) sampling technique, implemented through the emcee Python package~\cite{foreman2013emcee}.
\subsection{The observation of Cosmic chronometer dataset}\label{sec:4.1}
\begin{center}
	\begin{figure}
		\includegraphics[width=15cm, height=7cm]{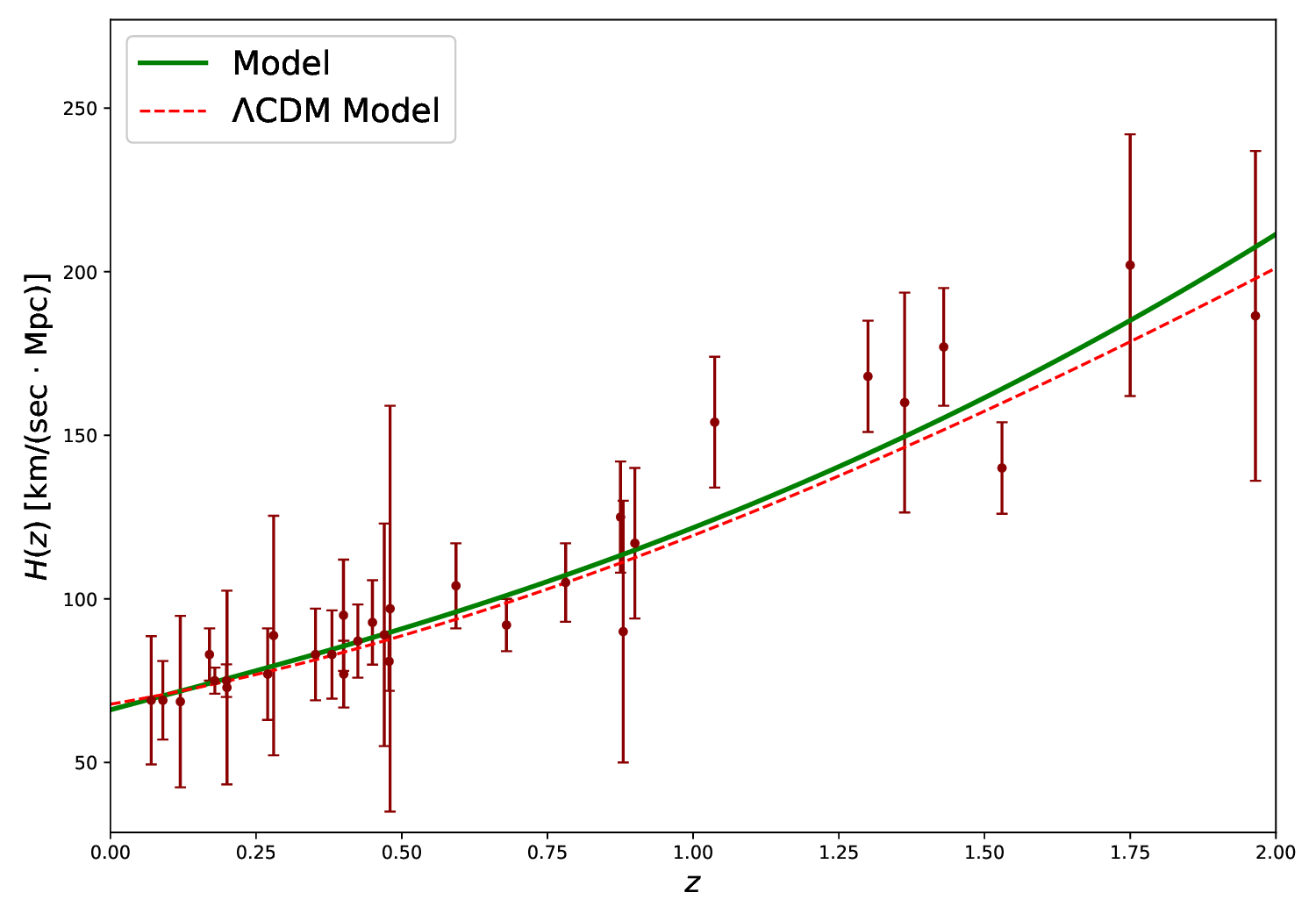}
		\caption{The best-fit $H(z)$ curve for the proposed model is compared with the $\Lambda CDM$ model.} 
		\label{fig:1}
	\end{figure}
\end{center}

To constrain the model parameters, we assess the observational viability of the proposed cosmological scenario using available data. The analysis utilizes a compilation of $31$ cosmic chronometer (CC) measurements~\cite{simon2005constraints,sharov2018predictions}, obtained via the differential age (DA) method applied to passively evolving galaxies over the redshift interval $0.07 \leq z \leq 1.965$ \cite{stern2010cosmic, moresco2015raising}. The primary objective is to estimate the best-fit values of the model parameters. Following the approach proposed by Jimenez and Loeb~\cite{jimenez2002constraining}, the Hubble parameter is related to redshift ($z$) and cosmic time ($t$) through $H(z) = \frac{-1}{(1+z)} \frac{dz}{dt}$. The free parameters $H_{0}$, $\omega_{0}$ and $n$ are constrained by minimizing the chi-squared $(\chi^{2})$ function, which is equivalent to maximizing the likelihood function~\cite{mandal2023cosmic}.
\begin{equation}{\label{25}}
	\chi^{2}_{CC}(\theta)=\sum_{i=1}^{31} \frac{[H_{th}(\theta,z_{i})-H_{obs}(z_{i})]^{2} }{ \sigma^{2}_{H(z_{i})}},  
\end{equation} 
\vspace{0.1cm}\\
where $H_{th}$ denotes the theoretical value of the Hubble parameter, $H_{obs}$ represents the observed value and $\sigma_{H}$ corresponds to the associated standard deviation of the observational data.
\vspace{0.1cm}\\
Figure $(\ref{fig:1})$ illustrates the error bars for the CC data points, together with the corresponding best-fit Hubble parameter curve.
\subsection{The observation of Pantheon dataset}\label{sec:4.2}
\begin{center}
	\begin{figure}
		\includegraphics[width=19cm, height=19.5cm]{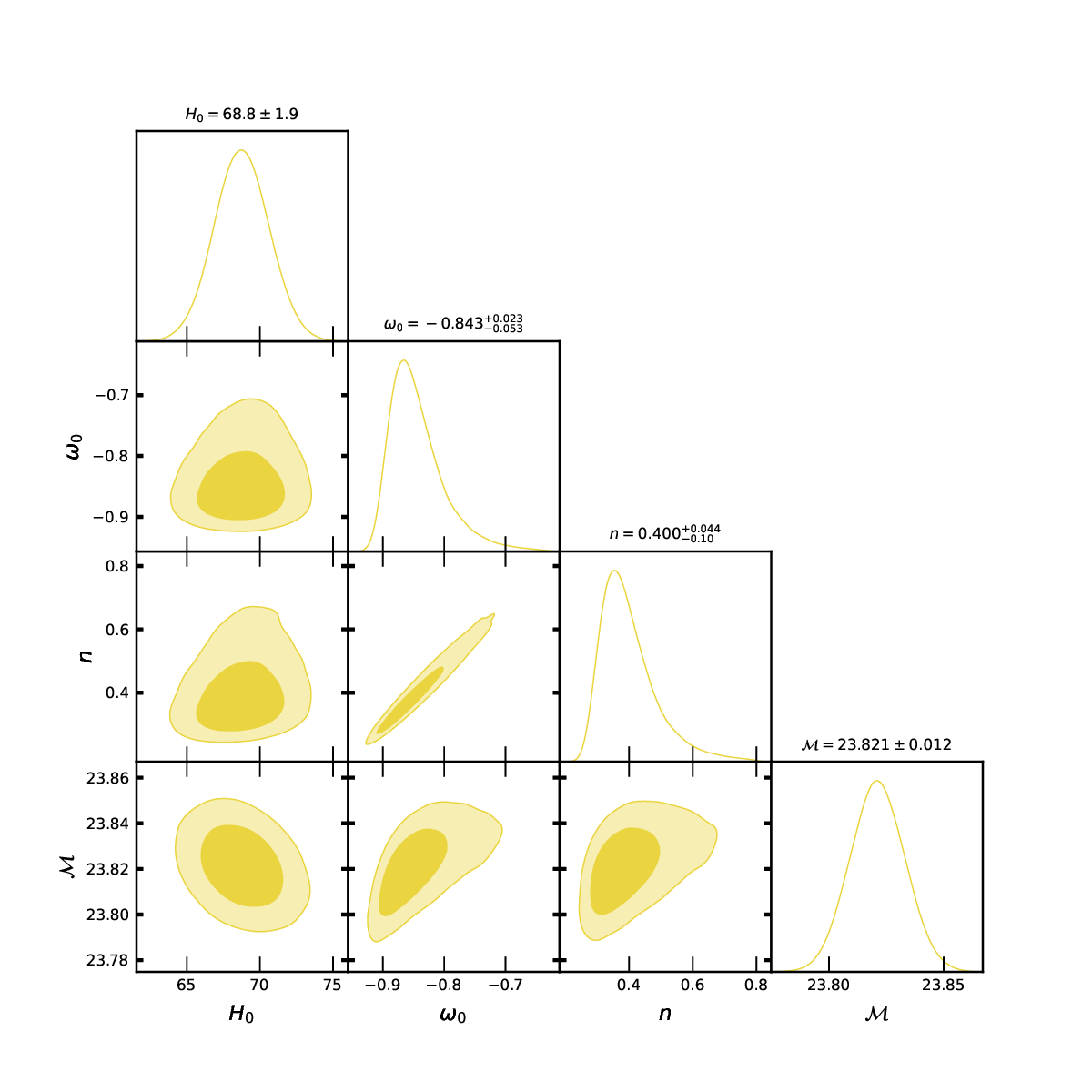}
		\caption{Marginalized $1D$ and $2D$ posterior contour map with median values of $H_{0}$, $\omega_{0}$ and $n$ using the Joint dataset.}
		\label{fig:2}
	\end{figure}
\end{center}

The present analysis makes use of the Pantheon compilation, which consists of 1048 Type Ia supernova (SNIa) data points covering the redshift range $0.01 < z < 2.26$~\cite{scolnic2018complete}. This comprehensive dataset combines observations from several major surveys, including the CfA1–CfA4 series~\cite{riess1999bvri,hicken2009improved}, the Pan-STARRS1 Medium Deep Survey~\cite{scolnic2018complete}, SDSS~\cite{sako2018data}, SNLS~\cite{guy2010supernova} and the Carnegie Supernova Project (CSP)~\cite{contreras2010carnegie}. In the MCMC analysis based on the Pantheon dataset, the theoretically predicted apparent magnitude $\mu_{th}(z)$ is expressed as
\begin{equation}{\label{26}}
\mu_{th}(z)=M+5log_{10}\left[\frac{d_{L}(z)}{Mpc}\right]+25,
\end{equation}
here $M$ represents the absolute magnitude, while the luminosity distance $d_{L}(z)$, having dimensions of length, is expressed as~\cite{odintsov2018cosmological}
\begin{equation}{\label{27}}
	d_{L}(z)=c(1+z)\int_{0}^{z}\frac{dz'}{H(z')}.
\end{equation}
In this context, $z$ represents the redshift of Type Ia supernovae (SNIa) measured in the cosmic microwave background (CMB) rest frame, while $c$ denotes the speed of light. The luminosity distance $d_L(z)$ is commonly rewritten in terms of a dimensionless, Hubble-independent quantity defined as $D_{L}(z) \equiv H_{0}d_{L}(z)/c$. Accordingly, equation~(\ref{26}) can be expressed in the form:
\begin{equation}{\label{28}}
	\mu_{th}(z)=M+5log_{10}\left[D_{L}(z)\right]+5log_{10}\left[\frac{c/H_{0}}{Mpc}\right]+25. 
\end{equation}
A degeneracy exists between $M$ and $H_{0}$ within the $\Lambda$CDM framework~\cite{ellis2012relativistic,asvesta2022observational}. To account for this, we introduce the combined parameter $\mathcal{M}$, defined as follows,
\begin{equation}{\label{29}}
	\mathcal{M}\equiv M+5log_{10} \left[\frac{c/H_{0}}{Mpc}\right]+25=M+42.38-5log_{10}(h), 
\end{equation}
taking $H_{0}=h \times 100$ $[\text{km}/(\text{sec}.\text{Mpc})]$, we perform the MCMC analysis by including these parameters along with the corresponding $\chi^{2}$ function for the Pantheon data, as defined in~\cite{asvesta2022observational}:
\begin{equation}{\label{30}}
	\chi^{2}_{P}= \nabla \mu_{i}C^{-1}_{ij}\nabla \mu_{j}.
\end{equation}
Here, $\nabla \mu_{i} = \mu_{obs}(z_{i}) - \mu_{th}(z_{i})$, where $C_{ij}^{-1}$ represents the inverse of the covariance matrix and $\mu_{th}$ is given by equation (\ref{28}). It is worth noting that the luminosity distance is directly governed by the evolution of the Hubble parameter. In this analysis, we utilize the emcee package~\cite{foreman2013emcee} in combination with the relevant theoretical framework to carry out maximum likelihood estimation (MLE) based on the joint (CC+Pantheon) dataset. The total chi-squared function employed for this purpose is defined as the sum $\chi^{2}_{CC} + \chi^{2}_{P}$. Figure~(\ref{fig:2}) illustrates the $1\sigma$ and $2\sigma$ confidence regions in the form of contour plots, together with the corresponding 1D posterior distributions obtained from the MCMC sampling of the CC+Pantheon joint datasets. The median estimates of the model parameters inferred from this analysis are listed in Table~(\ref{table:1}).
\begin{table}[htbp]
	\centering
	\renewcommand{\arraystretch}{3.5}  
	\fontsize{10pt}{10pt}\selectfont   
	\begin{tabular}{|c|c|c|c|c|c|c|c|c|c|}
			\hline
			Dataset & $H_{0}$[Km/(sec.Mpc)] & $\omega_{0}$ & $n$ & $\mathcal{M}$ & $q_{0}$ & $z_{t}$ & $t_{0}$(Gyr) \\
			\hline
			CC & $66.089^{+0.708}_{-0.710}$ & $-0.769^{+0.036}_{-0.030}$ & $0.477^{+0.046}_{-0.039}$ & - & $-0.2736$ & $0.726$ & $12.76$ \\
			\hline
			CC+Pantheon  & $68.8^{+1.9}_{-1.9}$  & $-0.843^{+0.023}_{-0.053}$ &  $0.400^{+0.044}_{-0.10}$ &  $23.821^{+0.012}_{-0.012}$ & $-0.4112$ & $0.795$ & $12.55$ \\
			\hline
	\end{tabular} 
     \caption{For both CC and joint datasets, the median values of the model parameters, along with the present values of $q_{0}$, $\omega_{0}$ and $t_{0}$.}
	\label{table:1}
\end{table}
\begin{center}
	\begin{figure}
		\includegraphics[width=15cm, height=7cm]{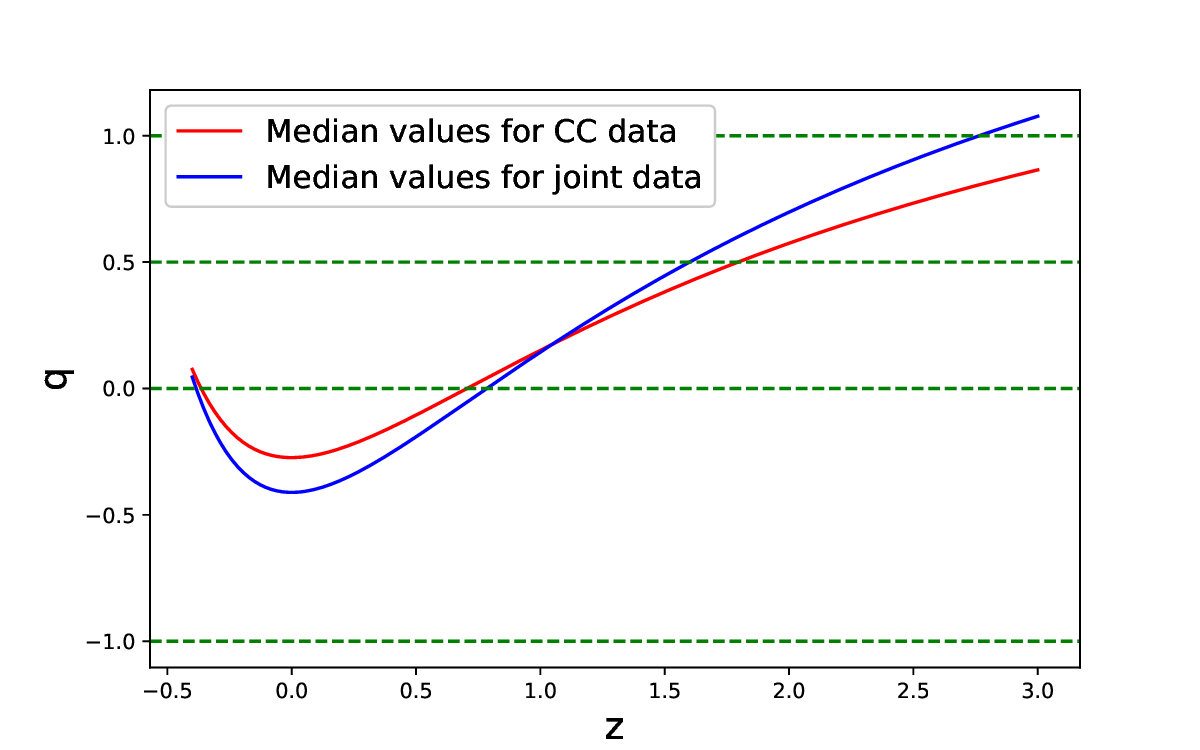}
		\caption{Plot of the deceleration parameter $(q)$ with $z$.} 
		\label{fig:3}
	\end{figure}
\end{center}
\section{The physical and dynamical characteristics of the model}\label{sec:5}
\subsection{Analysis of deceleration parameter}\label{subsec:5.1}
The deceleration parameter $q$ is a fundamental quantity for characterizing the expansion dynamics of the universe, as it provides insight into the nature of cosmic evolution. Different values of $q$ correspond to distinct expansion regimes: $q<0$ signifies an accelerating universe, while $q>0$ indicates a decelerating phase. In the regime $q<-1$, the universe undergoes super-accelerated expansion. Observational evidence suggests that specific values of the deceleration parameter are associated with different cosmic epochs, with $q=-1$, $q=\frac{1}{2}$ and $q=1$, corresponding to the de Sitter phase, matter-dominated era and radiation-dominated era, respectively. Accordingly, the general expression for the deceleration parameter is given by
\begin{equation}{\label{31}}
	q= -1+\frac{d}{dt}\frac{1}{H} .
\end{equation}
By substituting Eq.~(\ref{24}) into Eq.~(\ref{31}), we obtain
\begin{equation}
q(z)=-1+\frac{3}{2n(1+z)}\left[z+1+\omega_{0}\exp\left(\frac{z}{1+z}\right)\right].
\end{equation}
\vspace{0.01cm}\\
The evolution of the deceleration parameter is depicted in Fig.~(\ref{fig:3}). The figure clearly illustrates the transition of the universe from an earlier decelerated phase to the currently observed accelerated expansion. The present-day values of the deceleration parameter are obtained as $q_{0}=-0.2736$ for the CC dataset and $q_{0}=-0.4112$ for the joint dataset. The negative values of $q_{0}$ provide clear evidence for the ongoing accelerated expansion of the universe, in agreement with current observational constraints. The transition redshift, corresponding to the epoch at which the universe evolves from deceleration to acceleration, is found to be $z=0.726$ for the CC dataset and $z=0.795$ for the joint dataset. Overall, these results demonstrate that the proposed model successfully accounts for the present-day accelerated expansion, in accordance with recent observational findings. Interestingly, the model further indicates a future transition back to a decelerated expansion phase. This feature may be interpreted as a consequence of a possible interaction between dark energy and dark matter, leading to an effective transfer of energy from dark energy to dark matter. Similar evolutionary behavior has been reported in several recent studies~\cite{chakraborty2014third, castillo2023exponential, escobal2024can}, supporting the plausibility of such a scenario.
\begin{figure}[!htb]
	\captionsetup{skip=0.4\baselineskip,size=footnotesize}
	\begin{minipage}{0.50\textwidth}
		\centering
		\includegraphics[width=8.6cm,height=7cm]{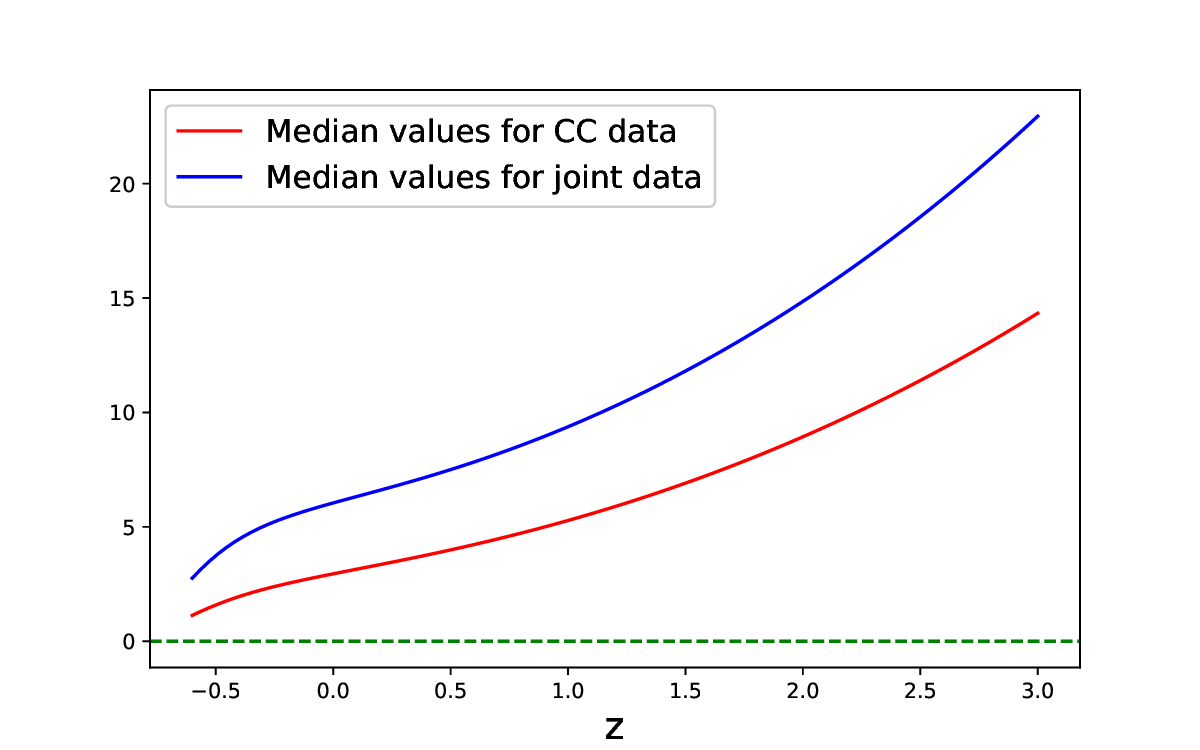}
		\caption{Plot of energy density ($\rho_{DE}$) with $\mathit{z}$.}
		\label{fig:4}
	\end{minipage}\hfill
	\begin{minipage}{0.50\textwidth}
		\centering
		\includegraphics[width=8.6cm,height=7cm]{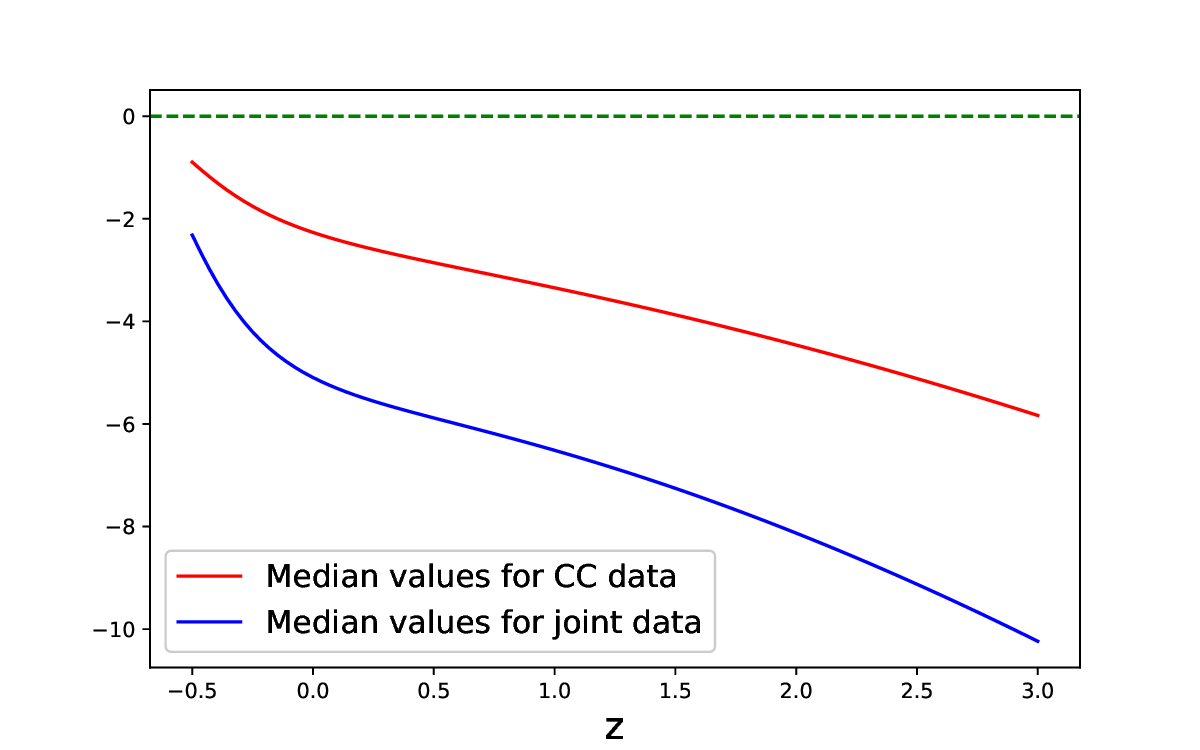}
		\caption{Plot of pressure ($p_{DE}$) with $\mathit{z}$.}
		\label{fig:5}
	\end{minipage}
\end{figure}
\subsection{Observational study of energy density, pressure and equation of state parameter
}\label{sec:5.2}
\begin{center}
	\begin{figure}
		\includegraphics[width=15cm, height=7cm]{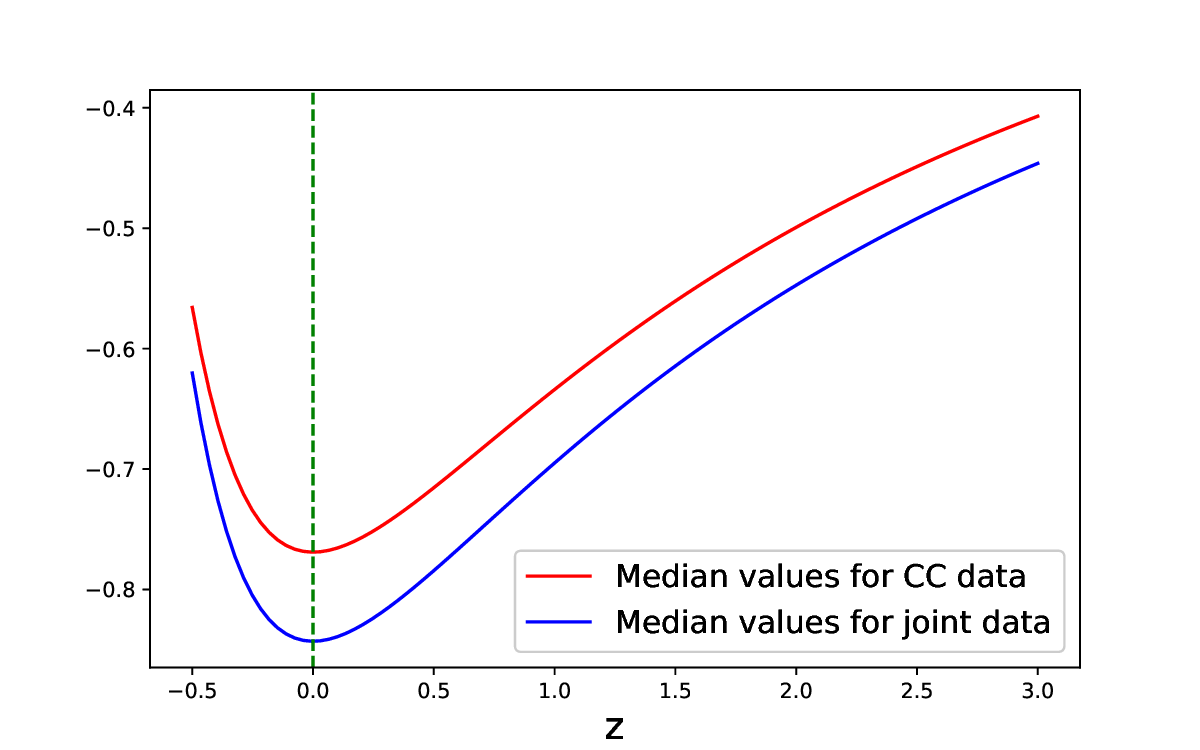}
		\caption{Plot of EoS parameter ($\omega_{DE}$) with $z$.}
		\label{fig:6}
	\end{figure}
\end{center}

We examine the physical properties of key cosmological quantities, namely the energy density and pressure. For the constrained model parameters, the energy density remains positive throughout the cosmic expansion, while the pressure exhibits negative values, driving the accelerated expansion of the universe. By using Eqs.~(\ref{17}), (\ref{18}) and (\ref{24}), the expressions for the DE density $(\rho_{DE})$ and DE pressure $(p_{DE})$ are obtained as,  
\begin{equation}{\label{33}}
\rho_{DE}(z) = \frac{(2n-1)\alpha 6^{n}}{2}\left(H_{0}^{2n} (1+z)^{3} \exp\left(3\omega_{0}\left[\left(\exp\left(\frac{z}{1+z}\right)\right)-1\right]\right)\right),
\end{equation}
\begin{equation} {\label{34}}
	\begin{split}
	p_{DE}(z)=& \left[\left(\frac{\alpha 6^{n}}{2}-n\alpha 6^{n}\right)+\left(3\alpha 6^{n-1}+(n-1)\alpha 6^{n}\right) \left(1+\frac{\omega_{0}}{1+z}\exp\left(\frac{z}{1+z}\right)\right)\right] \\& \left(H_{0}^{2n} (1+z)^{3} \exp\left(3\omega_{0}\left[\left(\exp\left(\frac{z}{1+z}\right)\right)-1\right]\right)\right).
	\end{split}
\end{equation} 
\vspace{0.1cm}\\
For the constrained model parameters, the evolution of the energy density ($\rho_{DE}$) and pressure ($p_{DE}$) are presented in Figs.~(\ref{fig:4}) and (\ref{fig:5}), respectively. The energy density exhibits a monotonic increase with redshift $z$ (corresponding to a decrease with cosmic time $t$) and remains positive throughout the cosmic evolution, supporting the role of DE in driving the present accelerated expansion of the universe. In contrast, the pressure remains negative over the considered redshift range, thereby providing the effective repulsive mechanism responsible for the present (at $z=0$) accelerated expansion of the universe. However, at late times, the model predicts a transition to a decelerated phase. This behavior may be attributed to the evolution of the pressure toward less negative or possibly positive, values in the future, which reduces the repulsive effect and leads to a gradual slowing down of the expansion. Such a feature highlights the dynamical nature of the model, where the condition required for sustained acceleration is not maintained indefinitely. We take $\alpha$ = $-1$ for these graphical illustrations. In cosmological studies, the Equation of State (EoS) parameter serves as a fundamental quantity for understanding the nature of DE. It characterizes the ratio between the pressure and the energy density of the cosmic fluid and is defined as $\omega$ = $\frac{p}{\rho}$. The value of $\omega$ provides key insights into the various evolutionary stages of the universe. For instance, $\omega=0$ corresponds to pressureless matter (dust), $\omega=\frac{1}{3}$ represents the radiation-dominated epoch, and $\omega=-1$ is associated with vacuum energy in a de Sitter universe. A key requirement for cosmic acceleration is that the EoS parameter satisfies $\omega <$ $-\frac{1}{3}$. This range encompasses both the quintessence regime ($-1 < \omega < -\frac{1}{3}$) and the phantom regime ($\omega < -1 $), each describing different dynamical behaviors of DE.
  \\  \\
The graphical evolution of the EoS parameter for the proposed model is illustrated in Fig.~(\ref{fig:6}). The present-day values at $z=0$ are found to be $\omega_{DE}=-0.769$ for the CC dataset and $\omega_{DE}=-0.843$ for the joint dataset. These values indicate that the model exhibits a quintessence-like behavior of DE at the current epoch for both datasets. The evolution of $\omega_{DE}$ further suggests a dynamical nature of DE, which is responsible for the observed accelerated expansion of the universe in the present epoch. However, the model also points toward a possible transition to a decelerated phase in the future. This behavior may be associated with an effective interaction between dark matter and dark energy, leading to a redistribution of energy that influences the late-time cosmic dynamics. Such a feature makes the model particularly interesting from a physical perspective. 
\subsection{Energy conditions}\label{sec:5.3}
\begin{figure*}[!htb]
	\centering
	\setlength{\fboxsep}{11pt} 
	\setlength{\fboxrule}{0.8pt} 
	
	\fbox{ 
		\begin{minipage}{0.95\textwidth}
			\centering
			\vspace{5pt}
			
			\subfloat[Plot of NEC \label{fig:7a}]{
				\includegraphics[width=0.32\textwidth]{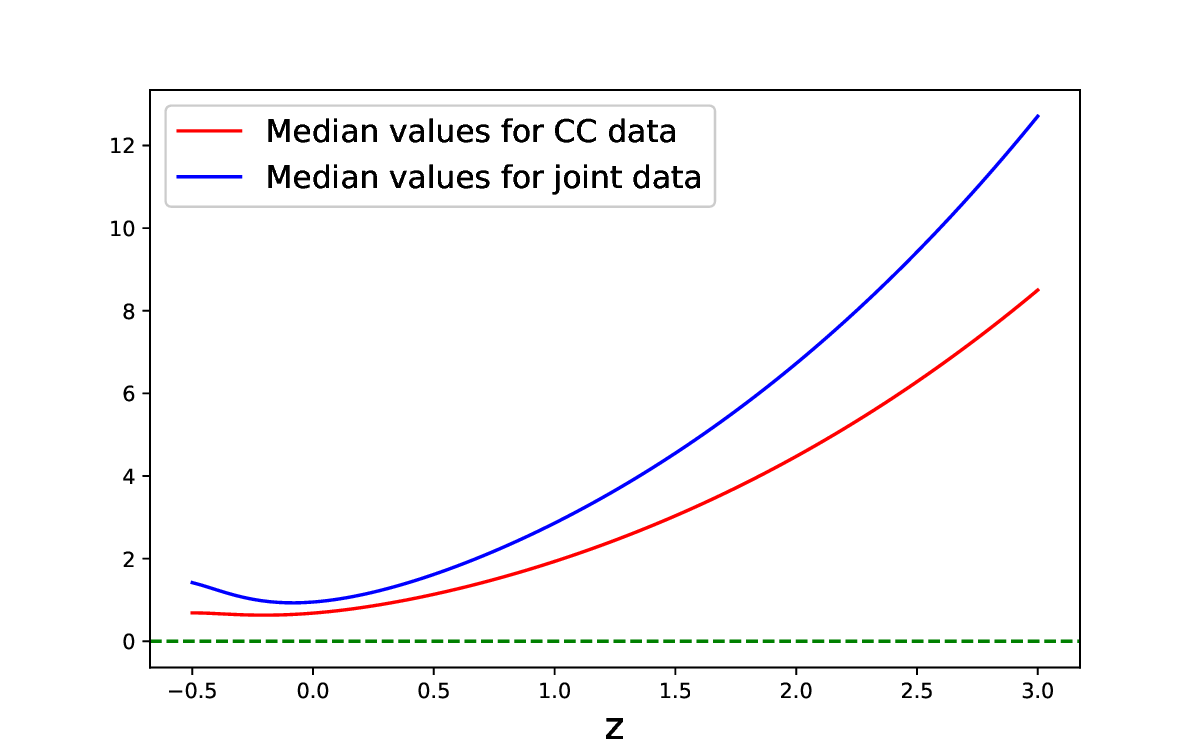}
			}
			\hfill
			\subfloat[Plot of DEC \label{fig:7b}]{
				\includegraphics[width=0.32\textwidth]{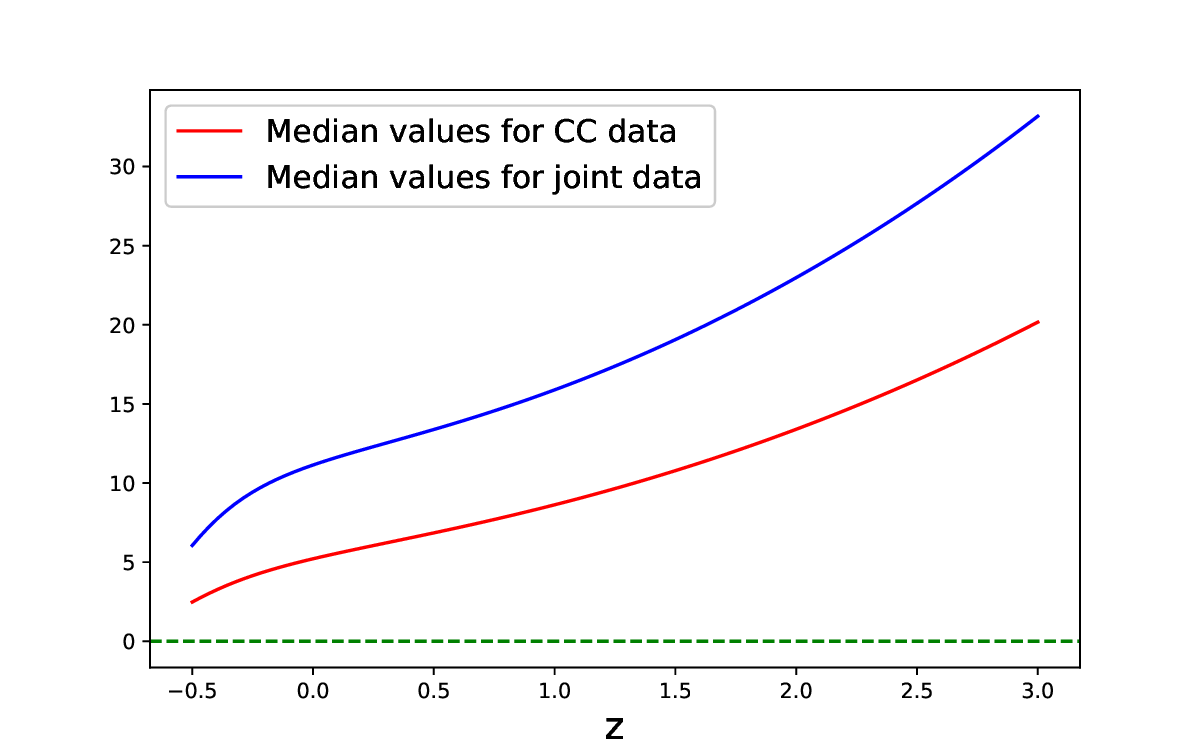}
			}
			\hfill
			\subfloat[Plot of SEC \label{fig:7c}]{
				\includegraphics[width=0.32\textwidth]{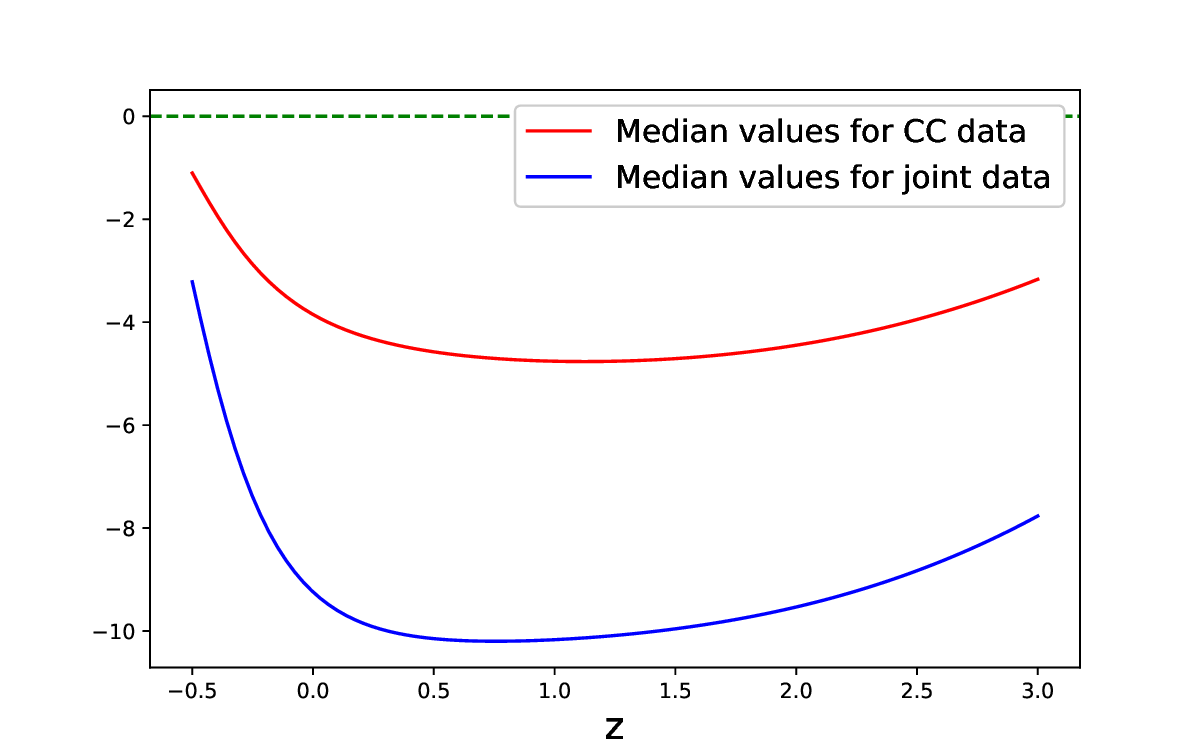}
			}
			
			\vspace{5pt}
		\end{minipage}
	} 
	
	\caption{The components of energy conditions with $z$.}
	\label{fig:7}
\end{figure*}

At a given point in spacetime, the point-wise energy conditions, which depend solely on the stress-energy tensor, are expressed as follows~\cite{visser1997energy,lalke2024cosmic, singh2022lagrangian}:
\begin{itemize}
    \item \textbf{NEC:-} Null energy condition requires $\rho_{eff} + p_{eff} \geq 0$, such that the effective energy density and pressure together remain non-negative.
	
    \item \textbf{WEC:-} Weak energy condition is satisfied when $\rho_{eff} \geq 0$ and  $ \rho_{eff} + p_{eff} \geq 0$, implying that the effective energy density and its sum with pressure are non-negative.
	
    \item \textbf{DEC:-} Dominant energy condition imposes that $ \rho_{eff}  \geq |p_{eff}| $, ensuring that the effective energy density is non-negative and exceeds the magnitude of the pressure.
	
	\item \textbf{SEC:-} Strong energy condition holds when $\rho_{eff} + p_{eff} \geq 0$ and $\rho_{eff} + 3p_{eff} \geq 0$, so that the effective energy density and pressure obey the necessary positivity conditions.
	
\end{itemize}
The Strong Energy Condition (SEC) is defined by the inequality $\rho_{eff} + 3p_{eff} \geq 0$, which plays a crucial role in the Raychaudhuri equation governing the acceleration or deceleration of cosmic expansion~\cite{mishra2025cosmological}. Observational evidence indicates that the SEC is violated at the present epoch, a feature that is essential for explaining the observed accelerated expansion of the universe. This violation implies that the effective pressure attains sufficiently negative values, pointing toward the dominance of a DE component. Consequently, an accelerating expansion arises when $\rho_{eff} + 3p_{eff} < 0$, reflecting the presence of a repulsive gravitational effect. It is worth noting that the SEC consists of two inequalities and the violation of either condition is sufficient to invalidate the SEC as a whole~\cite{mishra2025cosmological,myrzakulov2023quintessence}.
\vspace{0.2cm}\\
Figure~(\ref{fig:7}) illustrates the evolution of all energy conditions for the proposed model. The graphical analysis reveals that the Null Energy Condition (NEC), Weak Energy Condition (WEC) and Dominant Energy Condition (DEC) are consistently satisfied from the early universe up to the present epoch, thereby ensuring the physical admissibility of the model. In contrast, the SEC (specifically the constraint $\rho_{eff} + 3p_{eff} \geq 0$) is violated. This violation is a crucial requirement for driving the accelerated expansion of the universe and indicates the presence of an effective negative pressure component. The condition ($\rho_{eff} + 3p_{eff} < 0$) signifies the dominance of repulsive gravitational effects, which counteract the standard attractive nature of gravity at cosmological scales. Furthermore, the persistence of SEC violation during the present (at $z=0$) epoch highlights the sustained influence of DE in governing the current cosmic dynamics.
\\ \\
Remarkably, the model predicts that in the late-time future, the SEC may become satisfied ($\rho_{eff} + 3p_{eff} \geq 0$), signaling a transition back to a decelerated expansion phase. This behavior can be interpreted as a manifestation of a dynamical DE component or a possible energy transfer between dark energy and dark matter, leading to a gradual reduction of the repulsive effect. Such a feature highlights the rich dynamical evolution of the model, demonstrating its consistency with the present accelerated expansion and predicting a transition to a decelerated phase in the future cosmic evolution.
\begin{center}
	\begin{figure}
		\includegraphics[width=15cm, height=7cm]{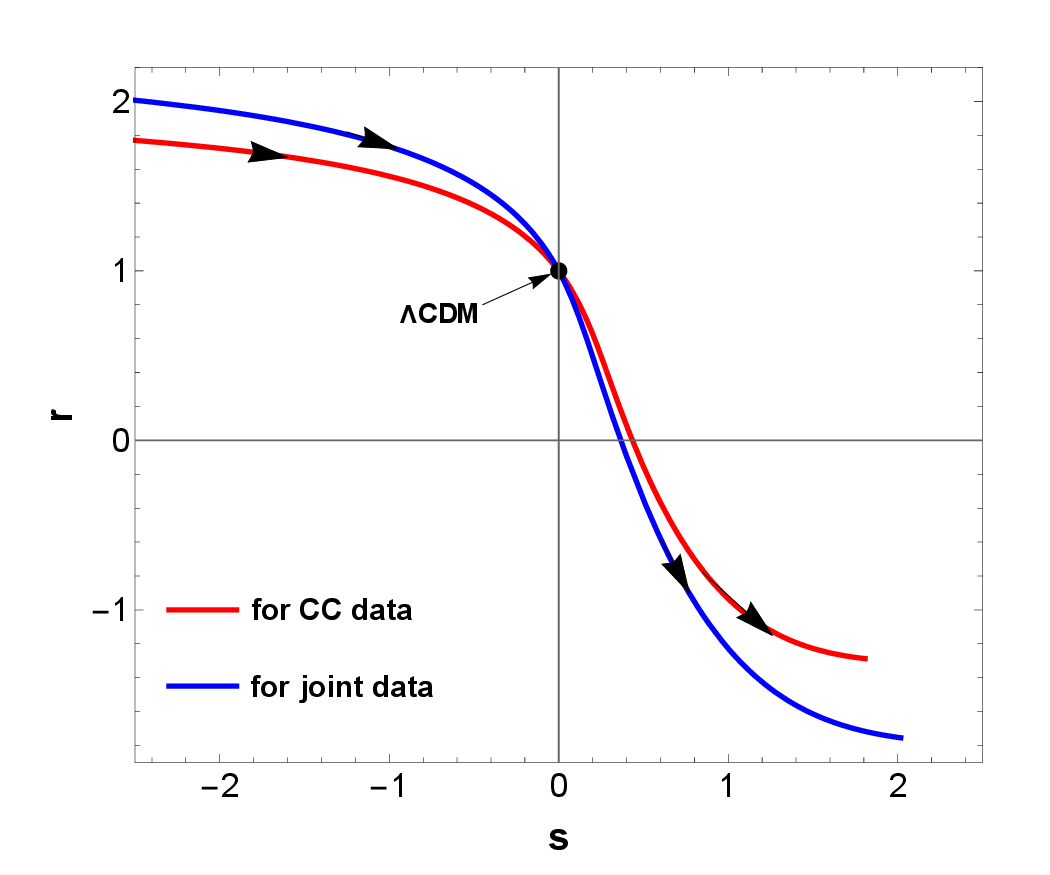}
		\caption{Plot of $s$ and $r$ plane.  }
		\label{fig:8}
	\end{figure}
\end{center}
\subsection{Statefinder diagnostic}\label{sec:5.4}
Geometric parameters play a pivotal role in characterizing the cosmological evolution of a model is well established. However, to rigorously explore alternative DE scenarios that diverge from the canonical $\Lambda$CDM framework, it is imperative to go beyond the conventional descriptors, namely the Hubble parameter ($H$) and the deceleration parameter ($q$). In this regard, higher-order derivatives of the scale factor $a(t)$ provide a more nuanced and sensitive probe, capable of revealing subtle dynamical distinctions among competing DE model. The statefinder diagnostic, expressed through the pair $\left\{r, s\right\}$~\cite{Sahni2003}, constitutes a powerful tool in this endeavor, offering a systematic means to classify and discriminate DE model based on their detailed evolutionary behavior. The statefinder parameters $\left\{r, s\right\}$ are defined as:
\begin{equation}{\label{41}}
r=\frac{\dddot a}{aH^{3}}, \quad s= \frac{r-1}{3(q-\frac{1}{2})}, \quad \text{where} \quad q\neq \frac{1}{2}.
\end{equation}
The evolution of various DE models discussed in the literature can be distinguished by their characteristic values of the statefinder pair $\left\{r, s\right\}$:
\begin{itemize}
	\item In the Chaplygin gas (CG) model, the parameters typically satisfy ($r >1$, $s <0$).
	
	\item The standard $\Lambda$CDM model is represented by $r = 1$ and $s = 0$.
	
	\item For Quintessence model, one generally finds ($r <1$, $s >0$).
	
	\item Holographic dark energy (HDE) model, ($r=1$, $s = \frac{2}{3}$).
	
	\item For Standard cold dark matter (SCDM), one may have ($r=1$, $s=1$).
	
\end{itemize}
The evolutionary trajectories of the proposed model in the $\left\{r, s\right\}$ diagnostic plane are illustrated in Figure~(\ref{fig:8}). The trajectory begins within the Chaplygin gas (CG) regime at early cosmic times, passes through the $\Lambda$CDM fixed point and subsequently evolves toward the quintessence region at the present epoch for both observational data estimates. Such behavior of the statefinder trajectories is consistent with trends reported in previous studies~\cite{fei2013statefinder}.
\subsection{Om($\mathit{z}$) diagnostics}\label{sec:5.5}
\begin{center}
	\begin{figure}
		\includegraphics[width=15cm, height=7cm]{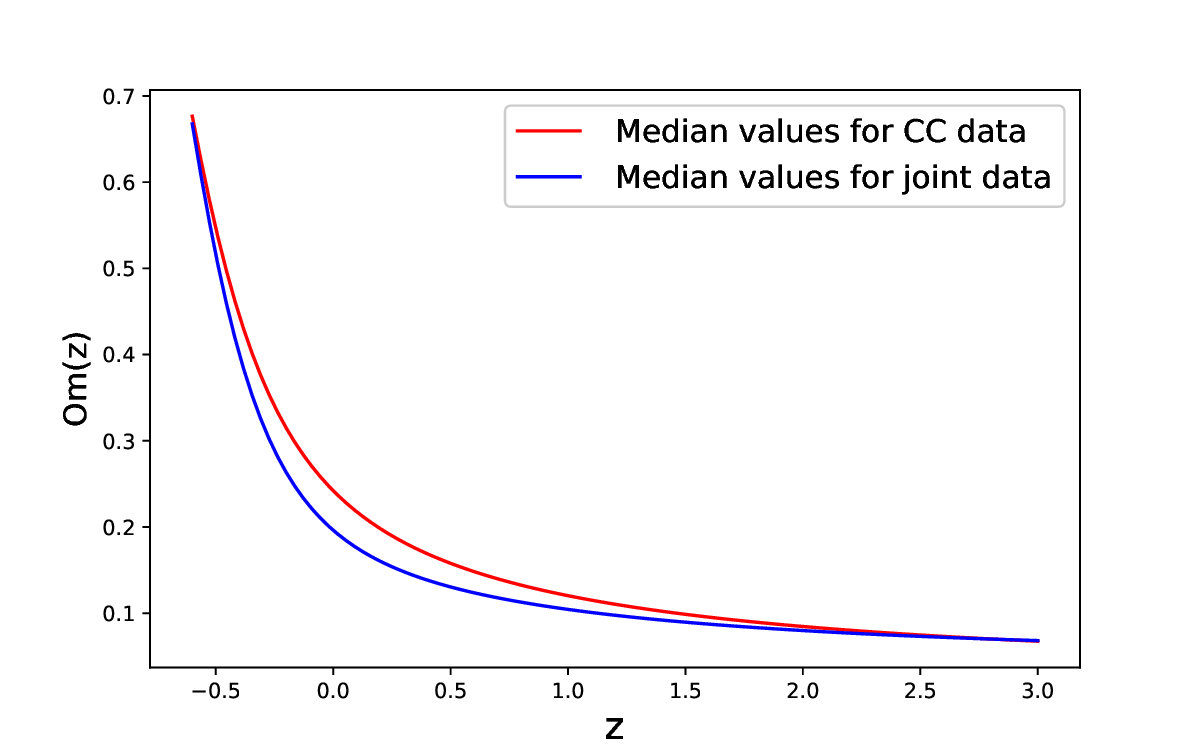}
		\caption{Plot of Om($\mathit{z}$) diagnostics with $z$.}
		\label{fig:9}
	\end{figure}
\end{center}

The Om($z$) diagnostic serves as a powerful tool for distinguishing between different dark energy models~\cite{sahni2008two}. For a spatially flat universe, it is defined as
\begin{equation} {\label{41}}
	Om(z) = \frac{\frac{H^{2}(z)}{H^{2}_{0}}-1}{(1+z)^{3}-1}.
\end{equation}
The slope of Om($z$) provides a direct probe of the dark energy nature: a positive slope signals phantom-like behavior, whereas a negative slope indicates quintessence-like behavior. A constant Om($z$) corresponds to the $\Lambda$CDM scenario. As shown in Fig.~(\ref{fig:9}), the evolution of Om($z$) for the proposed model clearly exhibits quintessence-like behavior at the present epoch ($z = 0$) across both observational datasets.
\subsection{The ($\omega_{\rm DE}-\omega'_{\rm DE}$) plane}\label{sec:5.6}
\begin{center}
	\begin{figure}
		\includegraphics[width=15cm, height=7cm]{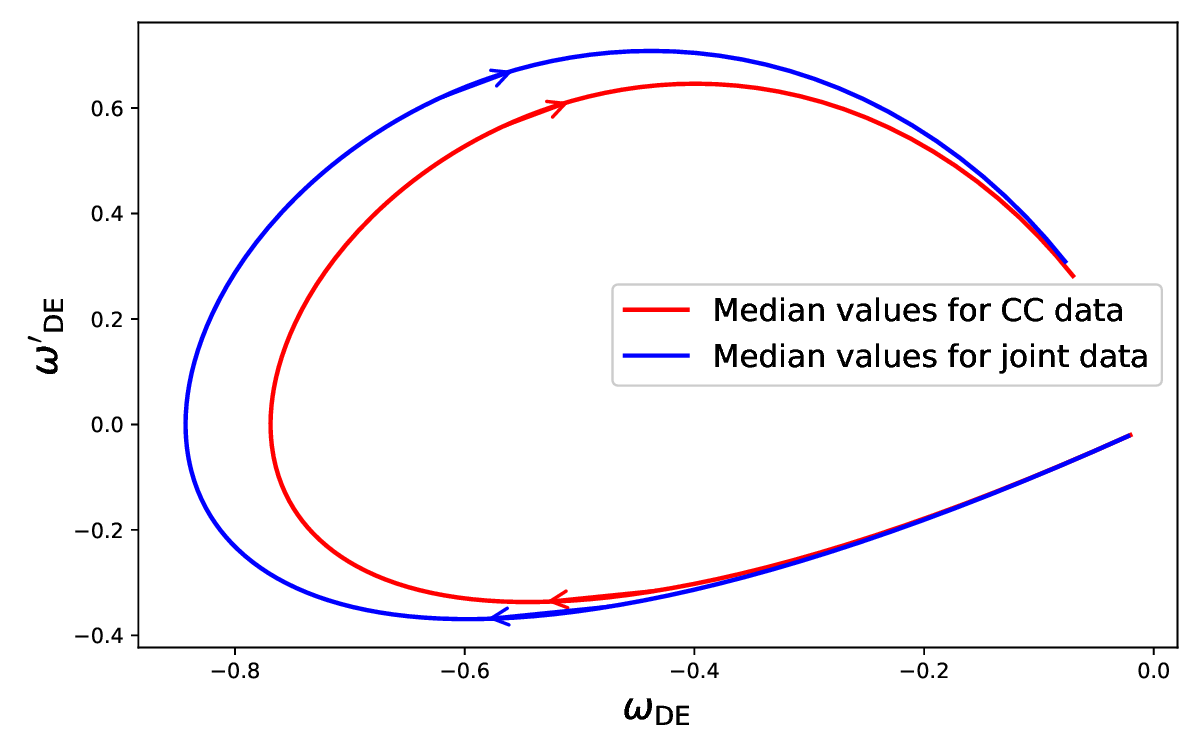}
		\caption{Plot of ($\omega_{\rm DE}-\omega'_{\rm DE}$) plane.  }
		\label{fig:10}
	\end{figure}
\end{center}

In this subsection, we perform a dynamical analysis of the DE EoS parameter ($\omega_{\rm DE}$) and its derivative with respect to $\ln a(t)$, denoted as ($\omega'_{\rm DE}=\frac{d\omega_{DE}}{d\ln(a)}$), for the proposed model. The ($\omega_{\rm DE}$–$\omega'_{\rm DE}$) plane serves as a powerful diagnostic tool for exploring dark energy dynamics, effectively delineating regions corresponding to accelerated cosmic expansion. This methodology, first proposed by Caldwell and Linder~\cite{caldwell2005limits} for quintessence scalar fields, offers a systematic approach for categorizing dark energy models according to their evolutionary trajectories. Two characteristic behaviors are distinguished on this plane: the thawing region ($\omega'_{\rm DE} > 0$ and $\omega_{\rm DE} < 0$) and the freezing region ($\omega'_{\rm DE} < 0$ and $\omega_{\rm DE} < 0$).
\vspace{0.2cm}\\
Figure~(\ref{fig:10}) shows the evolution of the ($\omega_{\rm DE}$–$\omega'_{\rm DE}$) plane for the proposed model, based on the median values of the model parameters from observational datasets. The trajectories form a closed-loop in the ($\omega_{\rm DE}$–$\omega'_{\rm DE}$) plane, representing the dynamical evolution of dark energy over cosmic time. At the present epoch ($z = 0$), the model resides in the freezing region ($\omega_{\rm DE} < 0$, $\omega'_{\rm DE} < 0$), indicating that dark energy behaves like quintessence while its EoS parameter is gradually stabilizing at the current epoch. In the future, $\omega_{\rm DE}$ increases toward zero, signaling a transition from the current accelerated expansion to a decelerated phase. The arrows along the trajectories indicate the forward direction of cosmic evolution. The similarity of the trajectories derived from both (CC and joint) datasets highlights the robustness of the predicted behavior. Overall, the ($\omega_{\rm DE}$–$\omega'_{\rm DE}$) plane provides a clear diagnostic that not only characterizes the current quintessence-like dynamics but also signals the onset of future deceleration, offering valuable insight into the long-term evolution of the universe.
\subsection{Thermodynamic analysis}\label{sec:5.7}
Thermodynamics shares a deep and intrinsic connection with gravitational theories that govern the large-scale dynamics of the universe. In a seminal work, Jacobson~\cite{jacobson1995thermodynamics} demonstrated that the Einstein field equations can be derived from the fundamental principles of black hole thermodynamics, thereby revealing the thermodynamic origin of spacetime dynamics. Subsequently, Padmanabhan~\cite{padmanabhan2002classical} showed that the gravitational field equations can be recast in the form of the first law of thermodynamics, further strengthening the profound interplay between gravity and thermodynamics. These pioneering contributions establish a compelling correspondence between gravitational dynamics and thermodynamic laws, providing a solid theoretical framework for exploring thermodynamic aspects of gravitational theories. Within this context, the generalized second law of thermodynamics (GSLT) plays a pivotal role and has been extensively investigated in Einstein gravity as well as in a wide class of modified theories of gravity~\cite{karami2012generalized,momeni2016generalized,mamon2021dynamics,pinki2023new}. The GSLT asserts that the total entropy of the universe defined as the sum of the entropy associated with the cosmological horizon and that of the matter-energy content enclosed within it, must evolve as a non-decreasing function of cosmic time. The validity of this law has also been examined in the framework of $f(T)$ teleparallel gravity in earlier studies~\cite{salako2013lambdacdm,karami2012generalized}, highlighting its significance in modified gravitational scenarios. In this work, we employ the Gong-Zhang parametrization of the DE EoS parameter within the framework of $f(T)$ teleparallel gravity. It is therefore of considerable interest to examine whether the proposed model satisfies the GSLT in this modified gravity setting. To this end, we carry out a detailed thermodynamic analysis of the model in the light of the GSLT. The total entropy of the universe can be expressed as
\begin{equation} {\label{43}}
	S_{tot} = S_{h} + S_{in},
\end{equation}
where $S_{tot}$ denotes total entropy, $S_{h}$ represents the horizon entropy and $S_{in}$ corresponds to the entropy of total fluid inside the horizon.
\vspace{0.2cm}\\ 
In order to assess the validity of the GSLT for the proposed model, we evaluate the time derivative of the total entropy $S_{tot}$. If the model satisfies the GSLT then we must observe $\dot{S}_{tot}$ $\geq$ 0. Using Eq.~(\ref{43}), we obtain the time variation of the total entropy as
\begin{equation} {\label{44}}
	\dot{S}_{tot} = \dot{S}_{h} + \dot{S}_{in}.
\end{equation}
Here, the overdot denotes differentiation with respect to cosmic time. To proceed further, we require explicit expressions for the horizon entropy $S_{h}$ and the entropy of the fluid enclosed within the horizon $S_{in}$. In the present analysis, we consider the entropy associated with the dynamical apparent horizon rather than that of the teleological event horizon, as the former is more suitable for describing the thermodynamic behavior of an evolving universe. For a dynamical apparent horizon, the entropy is given by $S_{h}=2\pi A$, where $A= 4\pi R_{h}^{2}$ represents the area of the apparent horizon. Here $R_{h}$ denotes the radius of the apparent horizon, for a spatially flat FLRW universe, is defined as $R_{h}= \frac{1}{H}$. Consequently, the entropy associated with the apparent horizon can be expressed as
\begin{equation} {\label{45}}
	S_{h} = \frac{8\pi^{2}}{H^{2}}
\end{equation}
and it’s rate of change is given by
\begin{equation} {\label{46}}
	\dot{S}_{h} = -16\pi^{2}\frac{\dot{H}}{H^{3}}.
\end{equation}
The Gibbs law of thermodynamics for fluid inside the horizon produces the relation
\begin{equation} {\label{47}}
	T_{in} dS_{in} = dE_{in}+ p_{t} dV_{h},
\end{equation}
where the subscript `$t$' denotes the total quantity and $V_{h}=\frac{4}{3} \pi R_{h}^{3}$ represents the volume enclosed by the apparent horizon. Therefore, the rate of change of entropy associated with the fluid inside the horizon can be written as
\begin{equation} {\label{48}}
	\dot{S}_{in} = \frac{(\rho_{t}+ p_{t})\dot{V}_{h}+ \dot{\rho}_{t} V_{h}}{T_{in}}.
\end{equation}
Under the assumption of thermal equilibrium between the cosmic fluid and the horizon, the temperature of the fluid inside the horizon, ($T_{in}$) is equal to the temperature of the dynamical apparent horizon ($T_{h}$)~\cite{duary2020brans}. This temperature is identified with the Hayward-Kodama temperature and can be expressed as
\begin{equation} {\label{49}}
	T_{h} = \frac{2H^{2}+\dot{H}}{4\pi H}.
\end{equation}
It is worth noting that, in the de Sitter spacetime where $\dot{H}=0$, this temperature reduces to the Hawking temperature, $T_{Hawking}= \frac{H}{2\pi}$~\cite{hawking1974black}. By employing Eqs.~(\ref{48}) and (\ref{49}), we obtain the rate of change of entropy of the fluid enclosed within the horizon as
\begin{equation} {\label{50}}
	\dot{S}_{in} = 16\pi^{2}\frac{\dot{H}}{H^{3}}\left(1+\frac{\dot{H}}{2H^{2}+\dot{H}}\right).
\end{equation}
Using Eqs.~(\ref{44}), (\ref{46}) and (\ref{50}), we derive the expression for the rate of change of the total entropy as
\begin{equation} {\label{51}}
	\dot{S}_{tot} = \frac{\left(\frac{4\pi \dot{H}}{H^{2}}\right)^{2}}{H\left(\frac{\dot{H}}{H^{2}}+2\right)}.
\end{equation}
\begin{center}
	\begin{figure}
		\includegraphics[width=15cm, height=7cm]{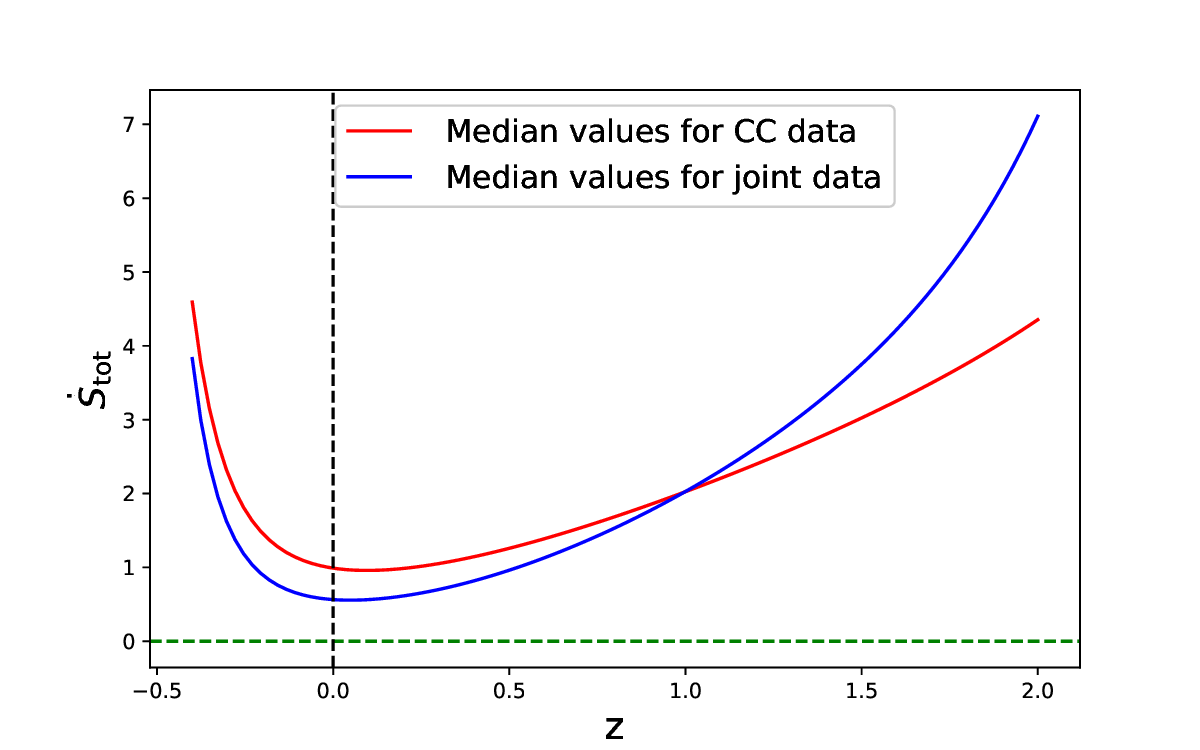}
		\caption{Plot of $\dot{S}_{tot}$ with $z$.  }
		\label{fig:11}
	\end{figure}
\end{center}
The validity of the GSLT requires that the time derivative of the total entropy satisfies $\dot{S}_{\rm tot} \geq 0$; otherwise, the model violates the GSLT. For the proposed model, this condition is fulfilled provided that $\frac{\dot{H}}{H^{2}} + 2 \geq 0$. Since the Hubble parameter $H$ remains positive, consistent with the observed accelerated expansion of the universe, this condition reduces to $\frac{\dot{H}}{H^{2}} \geq -2$. As the model is designed to describe the late-time dynamics of the universe, we analyze the behavior of $\dot{S}_{\rm tot}$ (after substituting $H(z)$ from Eq.~(\ref{24}) into Eq.~(\ref{51})) as a function of redshift $z$ over the interval $-0.5 \leq z \leq 2$, as shown in Fig.~(\ref{fig:11}). It is clearly observed that $\dot{S}_{\rm tot}$ remains positive throughout the entire redshift interval considered, including the present epoch ($z = 0$) and the future regime ($z < 0$). Therefore, the GSLT is consistently satisfied for both the CC and joint datasets within the considered framework, confirming the thermodynamic viability of the proposed model.
\subsection{Age of the universe}\label{sec:5.8}
The cosmic age $t(z)$, as a function of redshift $z$ for the cosmological model is expressed as~\cite{tong2009cosmic}
\begin{equation} {\label{42}}
	t(z) = \int_{z}^{\infty} \frac{dz}{(1+z)H(z)}.
\end{equation}
The present age of the universe ($t_{0}$), is determined by numerically evaluating this integral using the Hubble parameter $H(z)$ (from Eq.~(\ref{24})) at $z = 0$. For the proposed model, we obtain $t_{0} = 12.76$ Gyr from the CC dataset and $t_{0} = 12.55$ Gyr from the joint dataset. These results are consistent with observational estimates~\cite{Valcin2020}, supporting the model’s agreement with current cosmological data.
\section{Conclusions}\label{sec:6}
In this study, we investigate the cosmological evolution of the universe within the framework of $f(T)$ gravity in a spatially flat FLRW metric by considering a power-law form of the function $f(T)$, together with the Gong-Zhang parametrization of the DE EoS parameter, namely $\omega_{DE}(z)=\frac{\omega_{0}}{1+z}\exp\left(\frac{z}{1+z}\right)$. This parametrization enables a smooth and dynamical description of DE across different cosmic epochs. The model parameters $H_{0}$, $\omega_{0}$ and $n$ are constrained using the cosmic chronometer (CC) dataset along with the joint (CC+Pantheon) dataset through a Bayesian MCMC analysis and the corresponding results are summarized in Table~(\ref{table:1}). The best-fit curve of the Hubble parameter $H(z)$ demonstrate the compatibility of the model with the measured cosmic chronometer observations.
\vspace{0.2cm}\\
The dynamical analysis of the deceleration parameter shows that the model successfully captures the transition of the universe from an early decelerated phase to the currently observed accelerated expansion, as illustrated in Fig.~(\ref{fig:3}). The present-day values are $q_{0} = -0.2736$ (CC data) and $q_{0} = -0.4112$ (joint data), while the transition redshift is found to be $z_{t} = 0.726$ (CC data) and $z_{t} = 0.795$ (joint data). These results are consistent with current observational constraints and confirm the capability of the model to describe the expansion history of the universe with good accuracy.
\vspace{0.2cm}\\
A notable and distinctive prediction of the present scenario is that the currently accelerating universe does not continue to accelerate indefinitely and reflects the evolving nature of DE. Instead, the model predicts that the present phase of acceleration is a transient phenomenon and will eventually be followed by a future decelerated epoch. This behavior emerges naturally from the dynamical evolution of the EoS parameter and constitutes one of the most important outcomes of the present analysis. Unlike the standard $\Lambda$CDM cosmology, where the cosmological constant drives eternal acceleration, the present model suggests that the repulsive effect of DE gradually weakens with cosmic time. Consequently, the expansion of the universe may return to a decelerating regime in the future. Such a possibility provides a physically richer and more realistic description of cosmic evolution and may offer an alternative resolution to the problem of everlasting acceleration.
\vspace{0.2cm}\\
The evolution of the equation of state parameter further reinforces this interpretation, as depicted in Fig.~(\ref{fig:6}). In the very high-redshift regime ($z \gg 1$), the EoS parameter approaches $\omega_{DE} \rightarrow 0$, indicating an effective matter-like behavior consistent with the existence of a structure formation epoch. At the present epoch, the values $\omega_{DE} = -0.769$ (CC data) and $\omega_{DE} = -0.843$ (joint data) clearly place the model within the quintessence regime. 
In the asymptotic future, the EoS parameter again approaches zero, signaling a gradual decline in the dominance of DE and providing a natural mechanism for the eventual slowdown of cosmic expansion.
The physical behavior of the model is further validated through the evolution of the energy density and pressure for DE, as shown in Figs.~(\ref{fig:4}) and (\ref{fig:5}). The dark energy density remains positive throughout the cosmic evolution, ensuring physical consistency, while the pressure remains negative at the present epoch, driving the accelerated expansion. In the future, the pressure evolves toward less negative values, thereby reducing the repulsive effect and contributing to the transition toward deceleration.
\vspace{0.2cm}\\
The analysis of energy conditions (see Fig.~(\ref{fig:7})) reveals that the Null, Weak and Dominant Energy Conditions are upheld throughout the cosmic evolution, while the SEC is violated at the present epoch, a feature essential for accounting for the observed accelerated expansion. The model further indicates a prospective restoration of the SEC in the future, thereby signaling a transition toward a decelerated expansion phase. Additional support is provided by both geometrical and dynamical diagnostics, wherein the statefinder $\{r,s\}$, $Om(z)$ and the $(\omega_{DE}-\omega'_{DE})$ plane consistently favor a quintessence-like behavior at the present epoch with the model residing in the freezing region and the EoS parameter asymptotically approaching a stable configuration.
\vspace{0.2cm}\\
From a thermodynamic perspective, the model remains consistent with the generalized second law of thermodynamics, as the total entropy is observed to increase continuously with cosmic time. This confirms the thermodynamic viability of the proposed framework. Finally, the estimated age of the universe is obtained as $t_{0} = 12.76$ Gyr (CC) and $t_{0} = 12.55$ Gyr (joint), which are in reasonable agreement with observational estimates, further supporting the reliability of the model.
\vspace{0.2cm}\\
In summary, the present analysis demonstrates that the $f(T)$ gravity framework, characterized by a power-law form of $f(T)$ and the Gong-Zhang parametrization, provides a coherent, dynamically rich and observationally consistent description of cosmic evolution. Its ability to account for the present accelerated expansion while naturally predicting a future decelerated phase highlights the significance of dynamical DE model within modified gravity frameworks. 
This feature may offer a useful criterion for distinguishing such model from the standard cosmological paradigm in forthcoming high-precision observational surveys.
\section*{\textbf{Acknowledgements}}
One of the authers, G. P. Singh gratefully acknowledges the support provided by the Inter-University Centre for Astronomy and Astrophysics (IUCAA), Pune, India, under the Visiting Associateship Programme.

\end{document}